\documentclass[preprintnumbers,preprint,showpacs,english]{revtex4}
\usepackage{amsmath}
\usepackage{graphicx}
\usepackage{dcolumn}
\usepackage{bm}
\usepackage{epstopdf}

\begin{document}
\preprint{APS/123-QED}

\title{Effect of laser induced orbital momentum on magnetization switching}
\author{A. Kundu}
\affiliation{Department of Physics, University of Arizona, Tucson AZ 85721.}
\author{S. Zhang}
\affiliation{Department of Physics, University of Arizona, Tucson AZ 85721.}
\date{\today}

\pacs{71.70.Gm, 75.30.Et, 75.30.Hx, 75.70.-i, 85.75.-d}
%
%
%
\begin{abstract}
The observed magnetization switching by circularly polarized ultrafast laser pulses has been attributed to the inverse Faraday effect in which the induced non-equilibrium orbital momentum serves as an effective magnetic filed via spin-orbit coupling for magnetization rotation and switching. We critically examine this scenario by explicitly calculating the magnitude of the induced orbital momentum for generic itinerant band. We show that the calculated induced angular momentum is not large enough for reversing the magnetization in one laser pulse with the order of 100 femtosecond duration. Instead, we propose
that each laser pulse is capable to expand a reverse domain a few nano-meters and it takes multiple pulses to complete the magnetization reversal  process via domain wall motion.
\end{abstract}

\maketitle

\section{Introduction}
Laser-induced ultrafast magnetization switching have been experimentally demonstrated in a number of
rare earth and transition metal compounds or multilayers \cite{radu,stanciu,alebrand,mangin}. The essential role of the laser
is to heat the magnetic sample to a high temperature close to the Curie temperature in less than picosecond time scale and the subsequent cooling reorients the magnetization in the direction opposite to the initial one. When the laser field is unpolarized or linearly polarized, laser heating would reduce the magnitude of the magnetization known as demagnetization process \cite{beau}, but there is no external torque to rotate the magnetization away from its initial direction. The observed magnetization switching \cite{ostler} by laser heating with unpolarization laser beams must involve spatial/time reversal symmetry breaking processes such as highly non-uniform magnetization distributions. Model calculations have also indicated that the initial non-collinear canting angles between Fe and Gd moments in the ferrimagnetic compound FeGd are necessary to break the spatial inversion symmetry so that the internal exchange interaction upon the heating and cooling would favor one direction of the magnetization than the other \cite{chantrell,shufeng}. Thus, the heat induced switching is unlikely an intrinsic effect, namely, it depends on details of the inhomogeneous distribution of the magnetic configurations and it is doubtful that the controlled switching at nano-scales with nearly-single domain particles can be achieved.

Recently, all-optical helicity-dependent magnetization switching has been observed \cite{vahaplar, mangin,fullerton1} in which the switching occurs only for the laser pulse with a definitive helicity: a right-circularly polarized light (PL) is able to switch the magnetization from up to down while the left-circularly PL can switch it back, and the linearly-polarized or unpolarized lights do not do either switching. The circularly PL itself carries a definite angular momentum direction, and thus it provides the symmetry breaking 
needed for favoring one diection of magnetization over the other. The inverse Faraday effect (IFE), which is synonym with the laser induced non-equilibrium electron orbital moment, has been proposed as a leading mechanism \cite{review1,fesenko}.  If the laser induced orbital moment is sufficiently large and the spin-orbit coupling is sufficiently strong, the magnetization can be switched to the direction of the induced orbital moment that is determined by the polarization of the light. Indeed, theoretical models have already been developed for explaining the laser-induced magnetization
switching by the IFE \cite{lagoutte,hertel,pershan}. A formulation  in which the induced momentum is expressed in terms of exact band states have also been constructed \cite{oppeneer}. However, these calculations do not quantitatively
address whether the IFE alone is able to account for the experimental results. For example, it has been found experimentally that
a multiple laser pulses, of the order of a thousand, is required to achieve the magnetization reversal \cite{fullerton1}. Such accumulated reversal processes from multiple laser pulses have not been explained.

In this paper, we calculate the laser induced orbital momentum by considering several different models. The essential goal is to estimate the magnitude of the orbital momentum for a given value of the laser intensity used in the switching experiments. 
In Sec. II, we provide an order of magnitude estimates on the magnitude of the induced orbital moment based a free electron model. In Sec. III, we present our model calculation of the IFE for arbitrary itinerant bands, followed by a quantitative estimate of the induced orbital moments with experimental parameters in Sec. IV. We discuss possible mechanisms for observed magnetization switching in Sec. V and we conclude in Sec. VI.

\section{Simplified estimation of the induced orbital moment }

The idea of the magnetization switching by the IFE is that the induced non-equilibrium orbital moment generates an effective field on the spin via spin-orbit coupling
$H= \xi {\bf s}\cdot{\delta \bf L}$ where $\xi$ is the spin-orbit coupling strength, $\delta {\bf L}$ is the induced orbital moment. To switch the spin within the laser pulse duration $ t_p$, which is typically less than a pico-second, 
the effective field $H_{eff}= \xi \delta {\bf L} $ must be at least of the order of $1/(\gamma t_p) > 100$ (T) where $\gamma $ is the gyromagnetic ratio. With this simple reasoning, the IFE would require an induced orbital momentum at least on the order of $\delta L =0.1 \hbar$ per magnetic ion if we take the spin-orbit coupling strength of the order of 0.1  eV. To see whether the experimental parameters could generate such magnitude of the orbital momentum, we shall make two rough estimates below. 

First, we estimate the total angular momentum of
the circularly polarized light received by the magnetic film during laser radiation,
\begin{equation}
\delta {\bf L}_p = \pm \frac{c\epsilon_0 E_0^2 a^3t_p}{2d\omega} {\bf \hat{z}}
\end{equation}
where $\pm$ represents right and left circular polarizations of the laser, ${\bf \hat{z}}$ is the incident direction of the light, $c$, $\epsilon_0$, $E_0$, $t_P$, $d$, $\omega$ and $a$ are the speed of light, permeability, magnitude of electric filed, pulse duration, thickness of the sample, angular frequency and lattice constant respectively. By using the experimental values from Ref. \cite{mangin},
we find the average angular momentum of the laser absorbed by one atom is about $10^{-4} \hbar$ per each laser pulse, assuming that the angular momentum of the light is completely transferred to the electron orbitals and there is no relaxation during the process. Thus, the maximum angular momentum transferred from the light to the orbital moment is at least 2-3 orders of magnitude too small to account for the experimentally observed switching by a single laser pulse.

The second simple estimation is to consider the interaction of the light with a free electron. The classical equation of motion of the electron with the circularly polarized light is
\begin{equation}
m \frac{d^2 \bf r}{dt^2} = e  E_0 (\hat{\bf x} \cos \omega t  \pm \hat{\bf y}  \sin\omega t)
\end{equation}
where $E_0$ is the magnitude of the electric field, and we consider the normal incident  ($\hat{\bf z}$) of light
on the film. The above equation yields an orbital angular momentum of the electron,
\begin{equation}
\delta {\bf L}_c \equiv m {\bf r}\times (d{\bf r}/dt) = \pm  (eE_0)^2 /(m \omega^3)\hat{\mathbf{z}}.
\end{equation}
By using the experimental values of
$E_0$ and $\omega$ derived from Ref.~\cite{mangin}, we find that ${\delta \bf L}_c$ is about $10^{-4}-10^{-3} \hbar$ which is again 2-3 orders of magnitude too small to switch the magnetization. One might improve the estimation by replacing the free electron model with a bound electron
which is subject to its internal resonant frequencies and damping parameters \cite{zon}. With reasonable material parameters, the order of magnitude remains about the same.

Motivated by the failure of the above simple estimation to explain the experimental results, we consider below the IFE for general bands by using the quantum description of the electron orbitals.  
	
\section{Induced orbital moment} 

We use the time-dependent perturbation theory to calculate the induced orbital momentum of an arbitrary band state whose
wavefunction is denoted by
$\psi_{n \bf k}^{(0)} ({\bf r})$ where $n$ is the band index. The standard interaction
between the circularly-polarized light and the electron is
\begin{equation}
V(t)= -eE_0  \left(x \cos (\omega t-\frac{\omega z}{c})  \pm y \sin(\omega t-\frac{\omega z}{c})\right) \left[
\theta (t) - \theta (t-t_p) \right]
\end{equation}
where $\theta (t)$ is the step function, $t_p$ is the laser pulse duration and $x$, $y$, $z$ are three components of position operator (${\bf r}$). The first order correction to the wavefunction by the above perturbation is
\begin{equation}
\psi_{n \bf k}(t)=e^{-i \omega_{n \bf k} t} \psi_{n \bf k}^{(0)}+ \sum_{m{\bf k}'\neq n {\bf k} } c_{n{\bf k}, m{\bf k}'}(t)e^{-i \omega_{m{\bf k}'}t} \psi_{m {\bf k}'}^{(0)}
\end{equation}
where
\begin{equation}
c_{n{\bf k}, m{\bf k}'}(t) = -\frac{i}{\hbar}\int_0^{t}\mathrm{d}t' \int d^3{\bf r} \psi_{n \bf k}^{*(0)}V(t')\psi_{m {\bf k}'}^{(0)}e^{i \omega_{n{\bf k}, m {\bf k}'}t'}
\end{equation}
and $\omega_{n{\bf k}, m {\bf k}'}= \omega_{n{\bf k}}- \omega_{m {\bf k}'}$.

The time-dependent average z-component orbital momentum $L_z\equiv xp_y -yp_x$ can be
calculated by using the equation of motion, 
\begin{equation}
\frac{d}{d t}\left\langle  L_z \right\rangle_{n\bf k}=\left\langle \psi_{n\bf k} (t)\left|\left[L_z,V(t)\right] \right|\psi_{n \bf k} (t)\right\rangle
\end{equation}
To proceed futher, we explicitly use the Bloch states where $\psi_{n{\bf k}}^{(0)}({\bf r}) = u_{n\bf k} ({\bf r})
\exp(i{\bf k}\cdot {\bf r})$ for calculating the matrix elements. The spatial integration in Eq.~(6) can by carried out by separating the integration over a unit cell and  summation over all periodic unit cells, i.e., replacing $\int d^3 {\bf r} $ by $ \sum_{{\bf R}_i} \int_{cell} d^3 {\bf r}$, where ${\bf R}_i$ is the lattice site. The summation over ${\bf R}_i$ yields the crystal momentum conservation, ${\bf k}' = {\bf k} \pm (\omega /c) {\bf \hat{z}} + {\bf G}$, where
${\bf G}$ is the reciprocal lattice vector. For the integration within a unit cell, we take the slowly varying function $\exp[i(\omega/c)z] \approx 1$. After explicitly carrying out spatial and time integration, we find, 
\begin{equation}\label{eq:delL}
\delta L_z = \pm \frac{4e^2 E_0^2}{\hbar}\sum_{n,m,{\bf k},} \left|<u_{n\bf k}|\frac{\partial u_{m\bf k'}}{\partial k'_x} >
\right|^2 \frac{\sin^2\left[(\omega+\omega_{n{\bf k}}-\omega_{m{\bf k}'})t_p/2\right](f_{n{\bf k}}-f_{m{\bf k}'})}{(\omega+\omega_{n{\bf k}}-\omega_{m {\bf k}'})^2 + 1/\tau^2} .
\end{equation}
where $f_{m{\bf k}'}$ ($f_{n\bf k})$ are the Fermi distribution functions, ${\bf k}' = {\bf k}+(\omega/c){\bf \hat{z}}$, and $\tau $ is a phenomenological parameter representing the energy relaxation time (further discussion follows).

Comparing the above equation to that of the standard transition probability between two atomic energy levels by the light, one 
finds several differences. First, the Fermi  distribution function limits the transition between occupied and unoccupied states
of the bands. Second, the momentum conservation ${\bf k}'={\bf k} + (\omega/c) {\bf \hat{z}}$ implies that the difference between ${\bf k}'$ and ${\bf k}$ is small, thus the energy
difference would be small as well if the transition occurs between the same band ($n=m$); i.e., $\omega_{n\bf k} - \omega_{n {\bf k}'} 
\ll \hbar \omega$. Therefore, the major contribution of the orbital momentum comes from the interband 
transition ($n\neq m$) with nearly same ${\bf k} \approx {\bf k}'$ and $\omega_{n \bf k} - \omega_{m \bf k} \approx \pm \omega$.
Third, the lifetime $\tau$ of the excited states of the itinerant electrons is finite. Two essential contributions are impurity scattering
and  some other intrinsic scattering such as phonons and electron-electron interactions. A rough estimate of the relaxation
time would be shorter than $\hbar/k_BT$ (where $T$ is the temperature). If one takes the temperature at $T=500$ K, $\tau$ would be no longer than $10^{-14}$ s. We will assume the relaxation time $\tau$ as a parameter which is included in Eq.~(8). We point out,
the resonant condition $\omega_{n {\bf k}} - \omega_{{m \bf k}'} = \omega$ in Eq.~(8) would yield a singularly large contribution
with an infinite relaxation time. The role of the finite relaxation is to broaden the spectrum of the states contributing to the
induced orbital momentum.

\section{Estimation of induced orbital moment }

As an example of the application of Eq.~(8), we consider the transition of the simplified band structure depicted in Fig.~(1) where
two hypothetical bands are separated by a band gap $\hbar \omega_0$ with each band characterized by bandwidth $W_i$ ($i=1,2$). Both bands are parabolic. The matrix element in Eq.~(8)  depends on the details of the Bloch wavefunction. If we construct the Bloch state by a local atomic Wannier function $\phi_n({\bf r})$
\begin{equation}
u_{nk}(\mathbf{r})=\frac{1}{\sqrt{N}} \sum_{\mathbf{R}}e^{-i{\bf k} \cdot ({\bf r} - \mathbf{R})} \phi_n(\mathbf{r}-\mathbf{R})
\end{equation}
where N is the number of sites in the system, we find
\begin{equation}
< u_{n\bf k}|\frac{\partial u_{m\bf k'}}{\partial k'_x}>= \beta_{mn}+\sum_{\mathbf{\Delta}}\gamma_{mn}(\mathbf{\Delta})\cos(\mathbf{k}\cdot\mathbf{\Delta})
\end{equation}
where $\beta_{mn}=-\int\mathrm{d}\mathbf{r}~ \phi_m^*(\mathbf{r})x\phi_n(\mathbf{r})$ is the on-site dipolar matrix element,
and $\gamma_{mn}(\mathbf{\Delta})=-\int\mathrm{d}\mathbf{r}~ \phi_m^*(\mathbf{r}-\mathbf{\Delta})x\phi_n(\mathbf{r})$ is the  overlap integration of the wavefunctions between nearest neighbor atomic site ${\bf \Delta}$. In general, $|\beta_{mn}| \gg | \gamma_{mn}| $ if the Wannier orbital
of $\phi_n$ and $\phi_m$ do not have same spatial symmetry which would make $\beta_{mn}$ identically zero (known as the selection rule), and thus the matrix element would be weakly dependent on ${\bf k}$ and we simply take it as a constant. With the above simplifications, Eq.~(8) now reduces to,
\begin{equation}
\delta L_z = \pm C \left(\frac{a^2\omega^2}{4\pi^3}\right)\int_0^{\epsilon_F} \mathrm{d}  \epsilon g({\epsilon})  
\frac{\sin^2 \left[ (\omega-\omega_0 - \alpha \epsilon ) t_p/2 \right] }{ \left( \omega-\omega_0- \alpha \epsilon \right)^2 + 1/\tau^2}\equiv\pm C I
\end{equation}
where $C=2e^2 E_0^2a^2/({\hbar \omega^2})$, $g(\epsilon) $ is the density of states, $\alpha = (1+W_2/W_1)/\hbar $, and the
last equality defines the dimentionless qunatity $I$ to be numerically calculated.

To access the dependence of $\delta L_z$ on the parameters entering in Eq.~(11), we numerically compute $I$ for various plausible parameters relevant to the experimental materials. In Fig.~(2), we show the influence of the  (or the energy gap $\hbar\omega_0$) between occupied
and unoccupied band. Clearly, the maximum values occurs when $\omega - \omega_0$ is near the vicinity of $\epsilon_F$. More importantly, both the peak value and the peak width depend on the relaxation time. As we discussed earlier, the relaxation time limits the overall non-equilibrium orbital momentum: the laser induced excitations from the occupied states to the unoccupied states has been balanced by the relaxation processes.
Another important parameter is the laser pulse duration $t_p$. In Fig.~(3), we show the peak 
values derived from Fig.~2 as a function of the pulse width.  At a small $t_p$, the orbital momentum linearly increases with the pulse duration, and it saturates at a certain value, typically of the order of a few tens of fs. 
 \begin{figure}
 \centering
       
        \includegraphics[width=0.45 \textwidth]{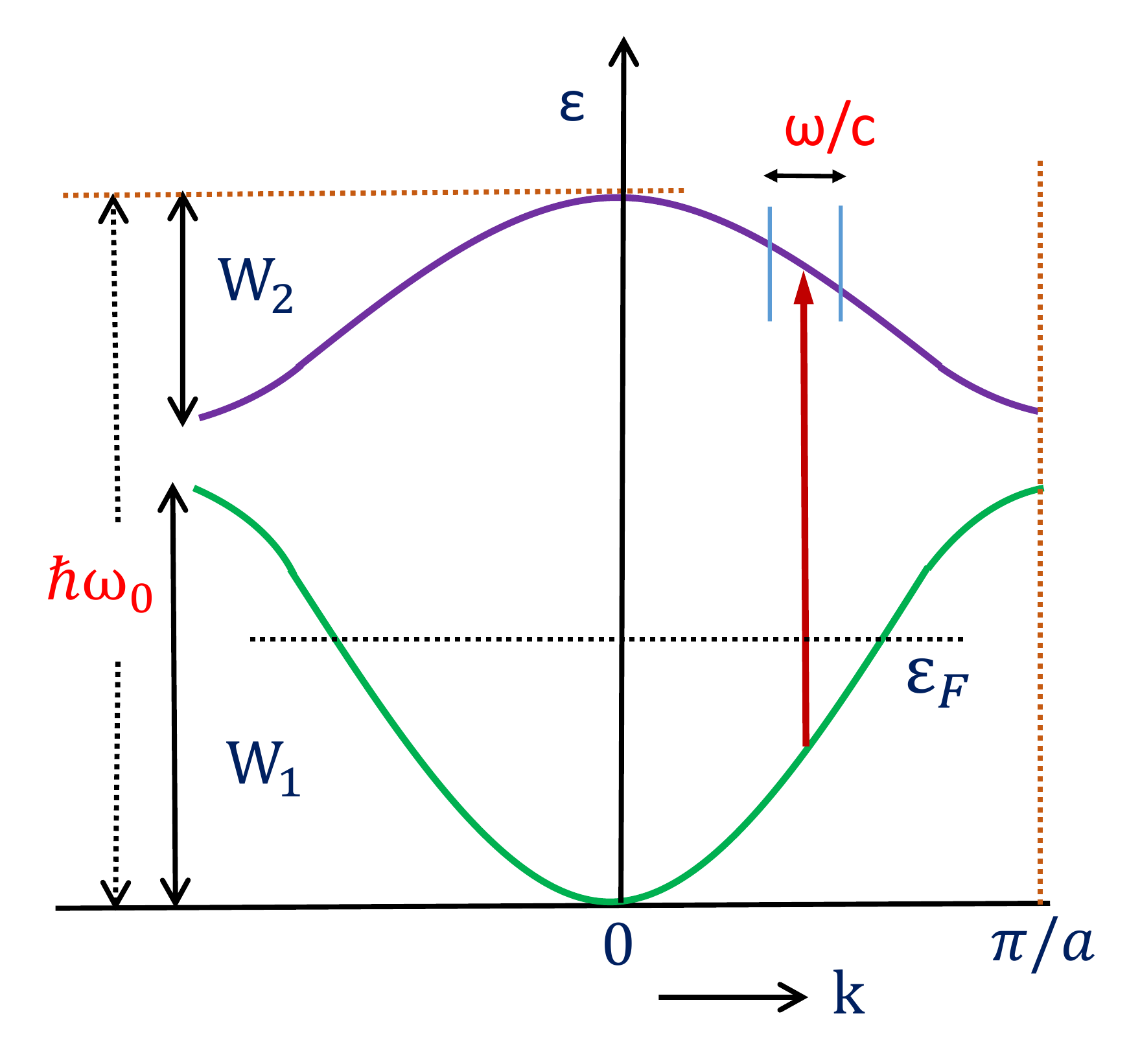}
                
                \label{fig1}
\caption{(color online) Laser induced transition between two hypothetical bands where $\epsilon_F$ represents Fermi energy and $W_1$, $W_2$ are the band width of lower and upper band respectively. The red arrows indicates the transition from the lower band at a particular $k$ state to the upper band at $k'-k \approx \omega/c$. }
\end{figure}

  \begin{figure}
 \centering
       
        \includegraphics[width=0.45 \textwidth]{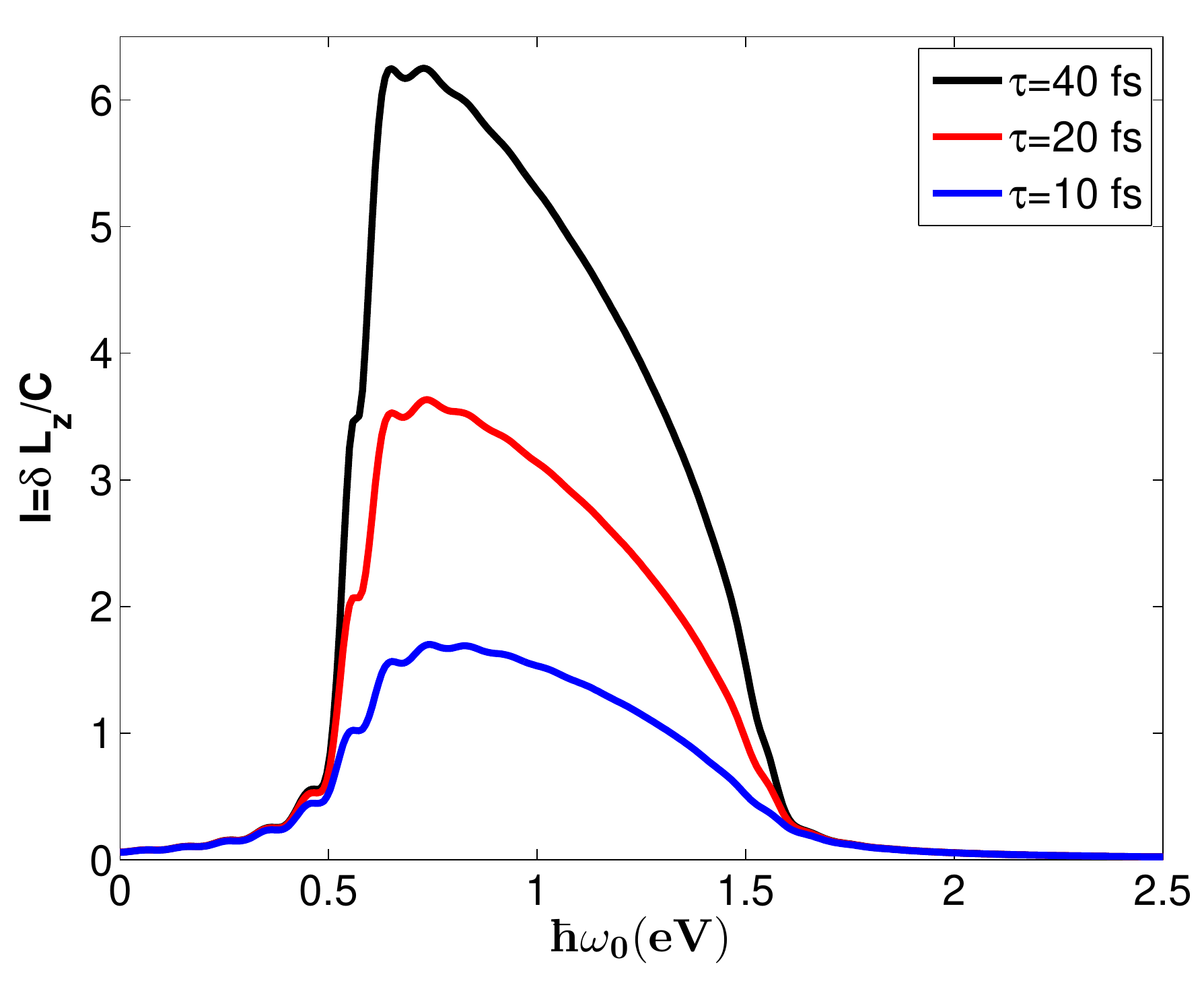}
                
                \label{fig1}
\caption{(color online) The induced orbital momentum as a function of the band separation $\hbar \omega_0$ for three different
relaxation times. The other parameters are $t_p = 50$ (fs) and $\alpha=1.5$ eV/$\hbar$.}
\end{figure}

 \begin{figure}
 \centering
       
        \includegraphics[width=0.45 \textwidth]{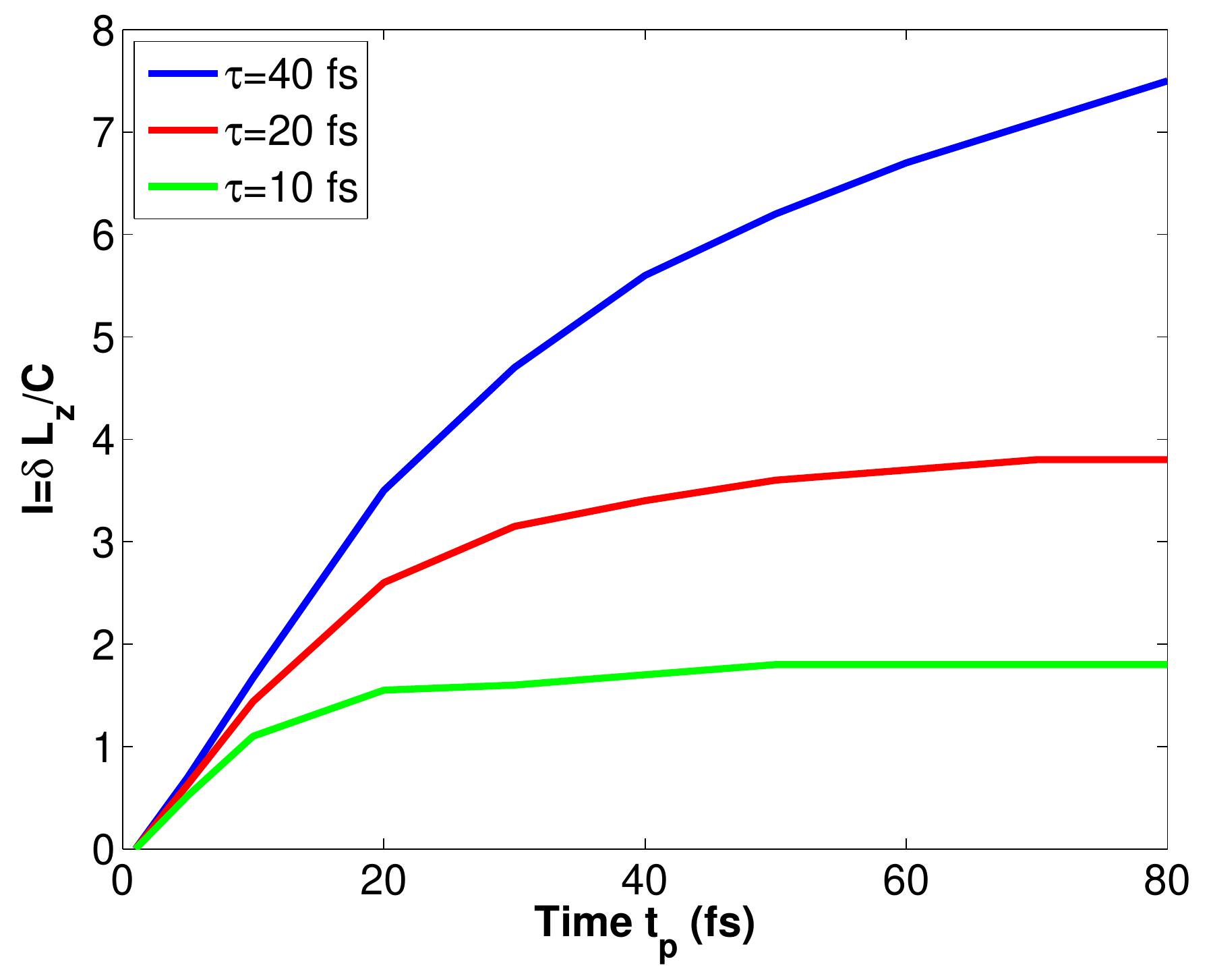}
                
                \label{fig3}
\caption{(color online) The induced orbital momentum as a function of laser pulse duration $t_p$ for $\alpha =1.5$ (eV/$\hbar$).}
\end{figure}

\section{Discussions}

We have calculated the laser induced orbital momentum for generic band structure. In experimental systems, band structures are far more complicated. We argue here that our simplified treatments of the band structure capture the essential features of the laser driven processes: since the most important
factor for the magnitude of the orbital momentum is the relaxation time as we have seen in Fig.~(2) and (3), the relaxation process is effectively 
washed out the detailed dispersion relation of the band. Since we have assumed a minimum relaxation rate in our estimation, the results
obtained here is an overestimate of the laser induced momentum.

It is interesting to compare the magnitude of the induced orbital momentum calculated here with two earlier estimations based on the
classical free electron model, Eq.~(3), and the total angular momentum transfer model, Eq.~(1). The relative ratios are
\[
\frac{\delta L_z}{\delta L^c} =2 \frac{ma^2\omega}{\hbar}I
\]
and 
\[ 
\frac{\delta L_z}{\delta L_p} =4\left(\frac{d}{a}\right)\left(\frac{e^2}{c\epsilon_0 a t_p \hbar \omega}\right)I
\]
By using the experimental values \cite{vahaplar, mangin}, all three estimates yield an angular momentum that are roughly same orders of magnitude, i.e., about $\delta L_z \approx 10^{-4}-10^{-3} \hbar$ per atom. We note that a recently published article  \cite{chen} claimed that theoretical estimation yields a reasonable magnitude to explain the above experiment; this error was due to mis-identification of the experimental value \cite{chen1}. Thus, the orbital momentum induced by single laser pulse is not sufficiently large to switch the magnetization of homogeneous magnetic materials.

Experimentally, the observed magnetization switching occurs when a large numbers of laser pulses, of the order of several thousand, are used \cite{fullerton1}. One may explain such accumulated switching processes as follows. The intense heating by laser pulses inevitably generates nucleation of domains. Each circularly polarized laser pulse creates an orbital momentum which provides
an effective field of the order of $1-10$T. While this transient field is unable to switch the domain, it can move the domain wall. If
we use a typical domain wall velocity of the order of 1-10 $\mu$m/ns, the domain wall could be displaced about $1-10$nm for a single laser pulse of $100$fs. Thus, several thousands of the laser pulses are capable to expand the nuclearation domain to $\mu m$ and
magnetization reversal is completed.

Finally, we wish to comment on the other possible mechanisms of the experimentally observed switching. Since the spatial coverage of the laser field is usually of the
order of several micrometers, the magnetization distribution would be highly non-uniform. Upon laser radiation, both heat and magnetization have spatial dependence. In a macroscopic sized film, the non-uniform distribution in turn leads to a strong and complicated dipolar interactions that govern the
magnetization dynamics and switching \cite{kronast}. In the earlier experiments, one can even observe the magnetization switching with an unpolarized laser beam and without any magnetic field \cite{ostler}; a result indicates the role of the non-uniform distribution in the symmetry-breaking processes.

\section{Conclusion and Acknowledgment}

Using the time-dependent quantum perturbation theory for generic itinerant bands, we explicitly calculate the orbital momentum of itinerant electrons induced by circularly polarized light. The magnitude of the induced orbital momentum is not sufficiently large to switch the magnetization in a single laser application. It is possible to achieve magnetization switching through domain wall nucleation and propagation by repeated applications of the laser pulses. Our result is consistent with the experimental observation. This work was supported by NSF (ECCS-1404542). 

\bibliographystyle{unsrt}
\bibliography{draft_01252017_bibtex}

\begin{thebibliography}{10}
\bibitem{chen1}
The author of Ref. [17] erranously quoted the laser intensity within the pulse
  width to be $500$ mJ/s cm$^2$ which is two orders of magnitude more than
  the~experimental values.
\end{thebibliography}

\end{document}